\documentclass[11pt]{article}

\def\noi{\noindent}

\def\barr{\left(\begin{array}}
\def\earr{\end{array}\right)}
\def\beq#1{\begin{equation}\label{#1}}
\def\eeq{\end{equation}}
\def\ber#1{\begin{eqnarray}\label{#1} \nqq}
\def\eer{\end{eqnarray}}

\newcommand{\bear}[1]{\begin{eqnarray}\label{#1}}
\newcommand{\ear}{\end{eqnarray}}

\catcode`\@=11 \@addtoreset{equation}{section}\catcode`\@=12
\newcommand{\N}{ {\bf N} }
\newcommand{\R}{ {\bf R }}

\newcommand{\sign}{\mathop{\rm sign}\nolimits}

\newcommand{\eps}{\varepsilon}
\newcommand{\tri}{\triangle}

\newcommand{\p}{\partial}
\newcommand{\nn}{\nonumber}

\newcommand{\fnm}{\footnotemark}
\newcommand{\fnt}{\footnotetext}

\begin{document}

\begin{flushright}        IGC-PFUR-08/2003  \end{flushright}

\vspace{15pt}

\begin{center}
\large\bf

 ON COMPOSITE S-BRANE SOLUTIONS \\
 WITH ORTHOGONAL INTERSECTION RULES

\vspace{15pt}

\normalsize\bf V.D. Ivashchuk\fnm[1]\fnt[1]{ivashchuk@mail.ru}

\vspace{5pt}

\it Center for Gravitation and Fundamental Metrology,
VNIIMS, 3/1 M. Ulyanovoy Str.,
Moscow 119313, Russia

\it Institute of Gravitation and Cosmology,
Peoples' Friendship University of Russia,
6 Miklukho-Maklaya St., Moscow 117198, Russia

\end{center}
\vspace{15pt}

\small\noi

\begin{abstract}

It is shown that non-composite intersecting $S$-brane  solutions
received in hep-th/0301095 are special case of cosmological-type
solutions with composite $p$-branes obtained in our earlier
publications.

\end{abstract}

\vspace{20cm}

\pagebreak

\normalsize

\section{Introduction}

This paper  is comment on a recent publication  \cite{Ohta}
devoted to intersection rules for $S$-branes. (The notion
``$S$-brane'' appeared in \cite{S1}, see also \cite{S2,Isbr} and
references therein.)

The  solution considered in \cite{Ohta} is not
a new one but:
  (i) for non-zero value of Robertson-Walker parameter
      $\sigma$ it is a special case of our solution from
 \cite{IMJ},
   (ii) for zero value of $\sigma$ it is a
        special case of the solutions  obtained
        earlier in \cite{Isbr}
  (see also \cite{IMtop}).

We note that the intersection rules considered in \cite{Ohta} are
not new ones but they special case of so-called ``orthogonal''
intersection rules from  \cite{IMJ}. (For earlier publications see
also \cite{IM11,AR,AIV,IMC} and refs. in \cite{Ohta}).

Here we deal with a model governed by the action
  \beq{1.1}
   S_g=\int d^Dx
   \sqrt{|g|}\biggl\{R[g]-h_{\alpha\beta}g^{MN}\p_M\varphi^\alpha
   \p_N\varphi^\beta-\sum_{a\in\tri}\frac{\theta_a}{n_a!}
   \exp[2\lambda_a(\varphi)](F^a)^2\biggr\}
  \eeq
where $g=g_{MN}(x)dx^M\otimes dx^N$ is a metric,
 $\varphi=(\varphi^\alpha)\in\R^l$ is a vector of scalar fields,
 $(h_{\alpha\beta})$ is a  constant symmetric
non-degenerate $l\times l$ matrix $(l\in \N)$,
 $\theta_a=\pm1$,
$F^a =    dA^a
=  \frac{1}{n_a!} F^a_{M_1 \ldots M_{n_a}}
dz^{M_1} \wedge \ldots \wedge dz^{M_{n_a}}$
is a $n_a$-form ($n_a\ge1$), $\lambda_a$ is a
1-form on $\R^l$: $\lambda_a(\varphi)=\lambda_{\alpha a}\varphi^\alpha$,
$a\in\tri$, $\alpha=1,\dots,l$.
In (\ref{1.1})
we denote $|g| =   |\det (g_{MN})|$,
$(F^a)^2_g  =
F^a_{M_1 \ldots M_{n_a}} F^a_{N_1 \ldots N_{n_a}}
g^{M_1 N_1} \ldots g^{M_{n_a} N_{n_a}}$,
$a \in \tri$.
Here $\tri$ is some finite set.
For pseudo-Euclidean metric of signature $(-,+, \ldots,+)$
all $\theta_a = 1$.

The paper is organized as following. In Section 2 we overview
cosmological-type solutions with composite intersecting $p$-branes
from \cite{IMJ,Isbr} obeying to  ``orthogonal'' intersection
rules. Section 3 is devoted to scalar-vacuum parametrization of
$S$-brane solutions that in a certain special case appears in
\cite{Ohta}.


 \section{Cosmological-type solutions with composite
 intersecting $p$-branes}

 \subsection{Solutions with $n$ Ricci-flat spaces}

Let us consider a family of
solutions to field equations corresponding to the action
 (\ref{1.1}) and depending upon one variable $u$
 \cite{Isbr} (see also \cite{IK,IMtop}).

These solutions are defined on the manifold
  \beq{1.2}
  M =    (u_{-}, u_{+})  \times
  M_1  \times M_2 \times  \ldots \times M_{n},
  \eeq
where $(u_{-}, u_{+})$  is  an interval belonging to $\R$,
and have the following form
 \bear{1.3}
  g= \biggl(\prod_{s \in S} [f_s(u)]^{2 d(I_s) h_s/(D-2)} \biggr)
  \biggr\{ \exp(2c^0 u + 2 \bar c^0) w du \otimes du  + \\ \nn
  \sum_{i = 1}^{n} \Bigl(\prod_{s\in S}
  [f_s(u)]^{- 2 h_s  \delta_{i I_s} } \Bigr)
  \exp(2c^i u+ 2 \bar c^i) g^i \biggr\}, \\ \label{1.4}
  \exp(\varphi^\alpha) =
  \left( \prod_{s\in S} f_s^{h_s \chi_s \lambda_{a_s}^\alpha} \right)
  \exp(c^\alpha u + \bar c^\alpha), \\ \label{1.5}
  F^a= \sum_{s \in S} \delta^a_{a_s} {\cal F}^{s},
 \ear
 $\alpha=1,\dots,l$; $a \in \tri$.

In  (\ref{1.3})  $w = \pm 1$,
 $g^i=g_{m_i n_i}^i(y_i) dy_i^{m_i}\otimes dy_i^{n_i}$
is a Ricci-flat  metric on $M_{i}$, $i=  1,\ldots,n$,
  \beq{1.11}
   \delta_{iI}=  \sum_{j\in I} \delta_{ij}
  \eeq
is the indicator of $i$ belonging
to $I$: $\delta_{iI}=  1$ for $i\in I$ and $\delta_{iI}=  0$ otherwise.

The  $p$-brane  set  $S$ is by definition
  \beq{1.6}
  S=  S_e \sqcup S_m, \quad
  S_v=  \sqcup_{a\in\tri}\{a\}\times\{v\}\times\Omega_{a,v},
  \eeq
 $v=  e,m$ and $\Omega_{a,e}, \Omega_{a,m} \subset \Omega$,
where $\Omega =   \Omega(n)$  is the set of all non-empty
subsets of $\{ 1, \ldots,n \}$.
Here and in what follows $\sqcup$ means the union
of non-intersecting sets. Any $p$-brane index $s \in S$ has the form
  \beq{1.7}
   s =  (a_s,v_s, I_s),
  \eeq
where
 $a_s \in \tri$ is colour index, $v_s =  e,m$ is electro-magnetic
index and the set $I_s \in \Omega_{a_s,v_s}$ describes
the location of $p$-brane worldvolume.

The sets $S_e$ and $S_m$ define electric and magnetic
$p$-branes, correspondingly. In (\ref{1.4})
  \beq{1.8}
   \chi_s  =  +1, -1
  \eeq
for $s \in S_e, S_m$, respectively.
In (\ref{1.5})  forms
  \beq{1.9}
  {\cal F}^s= Q_s  f_{s}^{- 2} du \wedge\tau(I_s),
  \eeq
$s\in S_e$, correspond to electric $p$-branes and
forms
  \beq{1.10}
  {\cal F}^s= Q_s \tau(\bar I_s),
  \eeq
  $s \in S_m$,
correspond to magnetic $p$-branes; $Q_s \neq 0$, $s \in S$.
Here  and in what follows
  \beq{1.13a}
  \bar I \equiv I_0 \setminus I, \qquad I_0 = \{1,\ldots,n \}.
  \eeq

All manifolds $M_{i}$ are assumed to be oriented and
connected and  the volume $d_i$-forms
  \beq{1.12}
  \tau_i  \equiv \sqrt{|g^i(y_i)|}
  \ dy_i^{1} \wedge \ldots \wedge dy_i^{d_i},
  \eeq
and parameters
  \beq{1.12a}
   \varepsilon(i)  \equiv {\rm sign}( \det (g^i_{m_i n_i})) = \pm 1
  \eeq
are well-defined for all $i=  1,\ldots,n$.
Here $d_{i} =   {\rm dim} M_{i}$, $i =   1, \ldots, n$;
 $D =   1 + \sum_{i =   1}^{n} d_{i}$. For any
 set $I =   \{ i_1, \ldots, i_k \} \in \Omega$, $i_1 < \ldots < i_k$,
we denote
  \bear{1.13}
  \tau(I) \equiv \tau_{i_1}  \wedge \ldots \wedge \tau_{i_k},
  \\
  \label{1.15}
  d(I) \equiv   \sum_{i \in I} d_i, \\
  \label{1.15a}
  \varepsilon(I) \equiv \varepsilon(i_1) \ldots \varepsilon(i_k).
 \ear

The parameters  $h_s$ appearing in the solution satisfy the
relations
 \beq{1.16}
  h_s = (B_{s s})^{-1},
 \eeq
where
 \beq{1.17}
  B_{ss'} \equiv
   d(I_s\cap I_{s'})+\frac{d(I_s)d(I_{s'})}{2-D}+
  \chi_s\chi_{s'}\lambda_{\alpha a_s}\lambda_{\beta a_{s'}}
  h^{\alpha\beta},
 \eeq
 $s, s' \in S$, with $(h^{\alpha\beta})=(h_{\alpha\beta})^{-1}$.

Here we assume that
 \beq{1.17a}
 ({\bf i}) \qquad B_{ss} \neq 0,
 \eeq
for all $s \in S$, and
 \beq{1.18b}
 ({\bf ii}) \qquad B_{s s'} = 0,
 \eeq
for $s \neq s'$, i.e. canonical (orthogonal) intersection rules
are satisfied.

The moduli functions read
 \bear{1.4.5}
  f_s(u)=
  R_s \sinh(\sqrt{C_s}(u-u_s)), \;
  C_s>0, \; h_s \eps_s<0; \\ \label{1.4.7}
  R_s \sin(\sqrt{|C_s|}(u-u_s)), \;
  C_s<0, \; h_s\eps_s<0; \\ \label{1.4.8}
  R_s \cosh(\sqrt{C_s}(u-u_s)), \;
  C_s>0, \; h_s\eps_s>0; \\ \label{1.4.9}
  |Q_s||h_s|^{-1/2}(u-u_s), \; C_s=0, \; h_s\eps_s<0,
  \ear
where $R_s = |Q_s|| h_s C_s|^{-1/2}$,
 $C_s$, $u_s$  are constants, $s \in S$.

Here
  \beq{1.22}
   \eps_s=(-\eps[g])^{(1-\chi_s)/2}\eps(I_s) \theta_{a_s},
  \eeq
$s\in S$, $\eps[g]\equiv\sign(\det(g_{MN}))$. More explicitly
(\ref{1.22}) reads: $\eps_s=\eps(I_s) \theta_{a_s}$ for
 $v_s = e$ and $\eps_s=-\eps[g] \eps(I_s) \theta_{a_s}$  for
 $v_s = m$.

Vectors $c=(c^A)= (c^i, c^\alpha)$ and
 $\bar c=(\bar c^A)$ obey the following constraints
 \beq{1.27}
  \sum_{i \in I_s}d_ic^i-\chi_s\lambda_{a_s\alpha}c^\alpha=0,
  \qquad
  \sum_{i\in I_s}d_i\bar c^i-
  \chi_s\lambda_{a_s\alpha}\bar c^\alpha=0, \quad s \in S,
   \eeq
  \bear{1.30aa}
  c^0 = \sum_{j=1}^n d_j c^j,
  \qquad
  \bar  c^0 = \sum_{j=1}^n d_j \bar c^j,
  \\  \label{1.30a}
  \sum_{s \in S} C_s  h_s +
    h_{\alpha\beta}c^\alpha c^\beta+ \sum_{i=1}^n d_i(c^i)^2
  - \left(\sum_{i=1}^nd_ic^i\right)^2 = 0.
 \ear

Here we identify notations  for $g^{i}$  and  $\hat{g}^{i}$, where
 $\hat{g}^{i} = p_{i}^{*} g^{i}$ is the
pullback of the metric $g^{i}$  to the manifold  $M$ by the
canonical projection: $p_{i} : M \rightarrow  M_{i}$, $i = 1,
 \ldots, n$. An analogous agreement will be also kept for volume forms etc.

Due to (\ref{1.9}) and  (\ref{1.10}), the dimension of
$p$-brane worldvolume $d(I_s)$ is defined by
 \beq{1.16a}
  d(I_s)=  n_{a_s}-1, \quad d(I_s)=   D- n_{a_s} -1,
 \eeq
for $s \in S_e, S_m$, respectively.
For a $p$-brane we have $p =   p_s =   d(I_s)-1$.

 \subsection{Solutions with one curved Einstein space and $n-1$
 Ricci-flat spaces}

The cosmological solution with Ricci-flat spaces
may be also  modified to the following case:
 \beq{1.2.2}
  {\rm Ric}[g^1] = \xi_1 g^1, \
  \xi_1 \ne0, \qquad {\rm Ric}[g^i] = 0, \ i >1,
 \eeq
i.e. the first space $(M_1,g^1)$ is  Einstein space of non-zero
scalar curvature and
other  spaces $(M_i,g^i)$ are Ricci-flat and
 \beq{1.2.3}
 1 \notin I_s,  \quad  \forall s  \in S,
 \eeq
i.e. all ``brane'' sub-manifolds  do not  contain $M_1$.

In this case the exact solution may be obtained by a little modifications
of the solutions from the previous subsection.
The metric reads as follows \cite{IMJ}
 \bear{1.2.4}
  g= \biggl(\prod_{s \in S} [f_s(u)]^{2 d(I_s) h_s/(D-2)} \biggr)
  \biggl\{[f_1(u)]^{2d_1/(1-d_1)}\exp(2c^1u + 2 \bar c^1)
   \times  \quad
  \\ \nn
  \times[w du \otimes du+ f_1^2(u)g^1] +
  \sum_{i = 2}^{n} \Bigl(\prod_{s\in S}
  [f_s(u)]^{- 2 h_s  \delta_{i I_s} } \Bigr)
  \exp(2c^i u+ 2 \bar c^i) g^i\biggr\}.
 \ear
where
 \bear{1.2.5}
  f_1(u) =R \sinh(\sqrt{C_1}(u-u_1)), \ C_1>0, \ \xi_1 w>0;
  \\ \label{1.2.6}
  R \sin(\sqrt{|C_1|}(u-u_1)), \ C_1<0, \  \xi_1 w>0;
  \\ \label{1.2.7}
  R \cosh(\sqrt{C_1}(u-u_1)),  \ C_1>0, \ \xi_1w <0; \\ \label{1.2.8}
  \left|\xi_1(d_1-1)\right|^{1/2} (u-u_1), \ C_1=0,  \ \xi_1w>0,
  \ear
  $u_1$ and $C_1$ are constants, $R = |\xi_1(d_1-1)/C_1|^{1/2}$, and

  \beq{1.2.9}
   \sum_{s \in S} C_s  h_s +
   h_{\alpha\beta}c^\alpha c^\beta+ \sum_{i=1}^n d_i(c^i)^2
   - \left(\sum_{i=1}^nd_ic^i\right)^2 = C_1\frac{d_1}{d_1-1}.
  \eeq

Now, vectors $c=(c^A)$ and $\bar c=(\bar c^A)$ satisfy
also additional constraints
 \bear{1.2.10}
  c^1 = \sum_{j=1}^nd_jc^j, \qquad
  \bar c^1= \sum_{j=1}^nd_j\bar c^j.
 \ear

All other  relations from the previous subsection are unchanged.

 {\bf Restrictions on $p$-brane configurations.}
The solutions  presented above are valid if two
restrictions on the sets of composite $p$-branes are satisfied \cite{IK}.
These restrictions
guarantee  the block-diagonal form of the  energy-momentum tensor
and the existence of the sigma-model representation (without additional
constraints) \cite{IMC}.

The first restriction reads
 \beq{1.3.1a}
  {\bf (R1)} \quad d(I \cap J) \leq d(I)  - 2,
 \eeq
for any $I,J \in\Omega_{a,v}$, $a\in\tri$, $v= e,m$
(here $d(I) = d(J)$).

The second restriction is following one
 \beq{1.3.1b}
  {\bf (R2)} \quad d(I \cap J) \neq 0,
 \eeq
for $I\in\Omega_{a,e}$ and $J\in\Omega_{a,m}$, $a \in \tri$.

 \section{Scalar-vacuum parametrization of $p$-brane solutions}

Here we show that the solutions with $p$-branes may be written in
a parametrization that appears in the absence of $p$-branes (when
$Q_s = 0$).

 \subsection{Scalar-vacuum solutions}

Let us  consider the scalar vacuum case
when all charges $Q_s$ are zero.

 \subsubsection{Solution with Ricci-flat spaces}

The scalar vacuum analogue
of the solution from subsection 2.1 reads
 \bear{1.3n}
  g=
   \exp(2c^0 u + 2 \bar c^0) w du \otimes du  +
   \sum_{i = 1}^{n}
   \exp(2c^i u+ 2 \bar c^i) g^i,
  \\ \label{1.4n}
   \varphi^\alpha = c^\alpha u + \bar c^\alpha,
 \ear
 $\alpha=1,\dots,l$;
where  $c^0$ and $\bar c^0$ obey to  (\ref{1.30aa}) and
 \beq{1.30an}
  h_{\alpha\beta}c^\alpha c^\beta+ \sum_{i=1}^n d_i(c^i)^2
  - \left(\sum_{i=1}^nd_ic^i\right)^2 = 0.
 \eeq

A special cases of this Kasner-like solution
were considered in \cite{I} (without scalar fields)
and in \cite{BZ0} (with one scalar field).

\subsubsection{Solutions with one  Einstein space
of non-zero curvature}

The scalar vacuum analogue
of the solution from subsection 2.2 is the following
one
 \bear{1.2.4n}
  g= [f_1(u)]^{2d_1/(1-d_1)} \exp(2c^1u + 2 \bar c^1)
   \times \\ \nn
  [w du \otimes du+ f_1^2(u)g^1]
  +  \sum_{i = 2}^{n} \exp(2c^i u+ 2 \bar c^i) g^i, \\
  \varphi^\alpha = c^\alpha u + \bar c^\alpha \nn,
 \ear
 $\alpha=1,\dots,l$;
 where $f_1(u)$ is defined in (\ref{1.2.5})-(\ref{1.2.8}) and
 \beq{1.2.9n}
  h_{\alpha\beta}c^\alpha c^\beta+ \sum_{i=1}^n d_i(c^i)^2
  - \left(\sum_{i=1}^nd_ic^i\right)^2 = C_1\frac{d_1}{d_1-1}.
 \eeq
Here $c^1$ and $\bar c^1$ obey to linear constraints (\ref{1.2.10}).

Special cases of this  solution
were considered in \cite{I} (without scalar fields)
and in \cite{BZ1,IM10,BZ2} (for one scalar field).

\subsection{Minisuperspace-covariant notations}

Here we show that the constraint (\ref{1.2.9}) may be
rewritten in a scalar-vacuum form  (\ref{1.2.9n}).
To do this we use the following
minisuperspace-covariant relations
 \cite{IMJ,IMC}:
 \beq{2.1}
  (\bar{G}_{AB})=\barr{cc}
  G_{ij}& 0\\
  0& h_{\alpha\beta}
  \earr,
 \qquad
 (\bar G^{AB})=\left(\begin{array}{cc}
 G^{ij}&0\\
 0&h^{\alpha\beta}
 \end{array}\right)
  \eeq
 are, correspondingly,
a (truncated) target space metric and inverse to it,
where  (see \cite{IMZ})
 \beq{2.2}
   G_{ij}= d_i \delta_{ij} - d_i d_j, \qquad
   G^{ij}=\frac{\delta^{ij}}{d_i}+\frac1{2-D},
 \eeq
and
 \beq{2.3}
   U_A^s c^A =
   \sum_{i \in I_s} d_i c^i - \chi_s \lambda_{a_s \alpha} c^{\alpha},
   \quad
   (U_A^s) =  (d_i \delta_{iI_s}, -\chi_s \lambda_{a_s \alpha}),
 \eeq
are co-vectors, $s=(a_s,v_s,I_s) \in S$ and
 $(c^A)= (c^i, c^\alpha)$.

In what follows we use a scalar product \cite{IMC}
 \beq{2.4}
  (U,U')=\bar G^{AB} U_A U'_B,
 \eeq
for $U = (U_A), U' = (U'_A) \in \R^N$, $N = n + l$.

The scalar products  for vectors
 $U^s$  were calculated in \cite{IMC}
 \beq{2.7}
  (U^s,U^{s'})= B_{s s'},
 \eeq
where  $s=(a_s,v_s,I_s)$,
 $s'=(a_{s'},v_{s'},I_{s'})$ belong to $S$ and
 $B_{s s'}$ are defined in (\ref{1.17}).
Due to relations (\ref{1.18b}) $U^s$-vectors
are orthogonal, i.e.
 \beq{2.7a}
 (U^s,U^{s'})= 0,
 \eeq
for $s \neq s'$.

The linear and quadratic constraints
from (\ref{1.27}), (\ref{1.2.10})
and (\ref{1.2.9}),
respectively, read in minisuperspace-covariant
form as follows:
 \beq{2.8}
   U_A^s c^A = 0, \qquad U_A^s \bar{c}^A = 0,
   \eeq
  $s \in S$,
  \beq{2.9}
   U_A^1 c^A = 0, \qquad U_A^1 \bar{c}^A = 0,
  \eeq
and
 \beq{2.10}
  \sum_{s \in S} C_s  h_s +
  \bar G_{AB} c^A c^B = C_1 \frac{d_1}{d_1-1}.
 \eeq

In  (\ref{2.9})
 \beq{2.11}
  U_A^1 c^A= - c^1+ d_i c^i,
  \qquad (U_A^1)=(-\delta_i^1+d_i,0),
 \eeq
is a co-vector corresponding to curvature
term. It obeys
 \beq{2.12}
  (U^1,U^1)=\frac1{d_1}-1<0, \qquad (U^1,U^{s})=0
 \eeq
for all $s\in S$.  The last relation
 (\ref{2.12})  follows from  (\ref{1.2.3}).

\subsection{New parameters for $S$-branes}

Let us consider the special case of the solutions
from sect. 2.2 with $h_s \eps_s > 0$ for all $s\in S$ and $w = -1$.
When all  $\theta_a > 0$ and  all internal spaces are of Euclidean signature
(and hence all $\eps_s > 0$)
we are led to  $S$-brane solutions (with $h_s$)  \cite{Isbr}.

We get from
 (\ref{1.4.5})-(\ref{1.4.9}) that
in this case
 \beq{2.13}
  f_s = R_s \cosh(\sqrt{C_s}(u-u_s)),
 \eeq
 where $R_s = |Q_s|| h_s C_s|^{-1/2}$ and $C_s>0$ for all $s\in S$.

Let us introduce new parameters
 \bear{2.14}
   c^A_{new} \equiv c^A -
   \sum_{s \in S } h_s \sqrt{C_s} U^{sA},  \\
   \label{2.14a}
   \bar c^A_{new} \equiv \bar c^A -
   \sum_{s \in S } h_s \sqrt{C_s} U^{sA},
  \ear
where $U^{sA} = \bar G^{AB} U^s_{B}$.

In terms of new parameters
the relations (\ref{2.8})-(\ref{2.10})
read
 \bear{2.8a}
  U_A^s c^A_{new} =  U_A^s \bar{c}^A_{new} = - \sqrt{C_s},
  \\ \label{2.9a}
 U_A^1 c^A_{new} = U_A^1 \bar{c}^A_{new} = 0,
 \\ \label{2.10a}
 \bar G_{AB} c^A_{new} c^B_{new} = C_1 \frac{d_1}{d_1-1}.
 \ear

The last relation when  written in components
has a scalar-vacuum form  (\ref{1.2.9n}), or
using relations  (\ref{2.9a}) (see (\ref{2.11}))
we get
 \beq{2.15}
   h_{\alpha\beta}c_{new}^\alpha c_{new}^\beta
   + \sum_{i=2}^n d_i(c^i_{new})^2
   + \frac{1}{d_1 - 1}
   \left(\sum_{i=2}^nd_ic^i_{new}\right)^2 = C_1\frac{d_1}{d_1-1}.
 \eeq

 In the special case of non-composite $S$-brane solutions with one scalar
 field and  internal  one-dimensional spaces of Euclidean signature
($M_i$, $i > 1$)
 the  relation  (\ref{2.15})
 is coinciding (up to notations) with the formula (26) from Ohta's
 paper  \cite{Ohta}. In this  case
 relations  (\ref{2.13}), (\ref{2.8a})  are in agreement  with formulas
 (37) and (38), respectively,  from \cite{Ohta}.

 Thus,   in the special
 case under consideration
 when the factor space $M_1$ is a spherical
 or a hyperbolic  space  we are led to the solution from \cite{Ohta}.
 \fnm[2]\fnt[2]{There are also two extra parameters in
 the solution from \cite{Ohta} standing for re-scaling of time and
 $M_1$ manifolds,  respectively.}

 An analogous consideration may be carried out for solutions
 from sect. 2.1
 when $M_1$ is a flat space and $S$-branes do not ``live'' in $M_1$
 (see  section 3 from \cite{Isbr}). In this case one
 can also verify that Ohta's solution is a special case of
 our solution from sect. 2.1.

\section{Conclusions}

In this paper we over-viewed cosmological-type solutions with
composite intersecting $p$-branes from \cite{IMJ,Isbr} obeying to
``orthogonal'' intersection rules. We obtained a scalar-vacuum
parametrization of $S$-brane solutions that allow us to identify
recent solution obtained  in \cite{Ohta} as a special case of our
cosmological type (e.g. S-brane) solutions.

It should be noted that the method of derivation of solutions used
by N. Ohta in \cite{Ohta} is not coinciding with our approach
based on reduction to Toda-like systems \cite{IMJ,IMtop}. It
should be also noted that our approach gives a rather systematic
way to derivation of intersection rules using the integrability
conditions for Toda-like systems \cite{IMJ,IK,IMtop}. In this
approach intersection rules have a minisuperspace-covariant form,
i.e. they are formulated in terms of scalar products of  brane
$U$-vectors and are classified by Cartan matrices of (semi-simple)
Lie algebras. The intersection rules considered in this paper
correspond to the Lie algebra $A_1 + \ldots + A_1$.

\begin{center}
{\bf Acknowledgments}
\end{center}

 This work was supported in part by the Russian Ministry of
 Science and Technology, Russian Foundation for Basic Research
 (RFFI-01-02-17312-a) and Project DFG (436 RUS 113/678/0-1(R)).


\small
\begin{thebibliography}{99}

 \bibitem{Ohta}
 N. Ohta,  Intersection rules for S-branes,
 {\it Phys. Lett.}  {\bf B 558}, 213 (2003);
 hep-th/0301095

 \bibitem{S1}
 M. Gutperle and A. Strominger,  Spacelike branes, {\it JHEP}
 {\bf 0204}, 018 (2002); hep-th/0202210.

  \bibitem{S2}
  C.M. Chen, D.M. Gal'tsov and M. Gutperle,  S-brane
  solutions in supergravity theories;
  {\it Phys. Rev.}  {\bf D 66}, 024043 (2002);
  hep-th/0204071.

 \bibitem{IMJ}
  V.D. Ivashchuk and V.N. Melnikov,
  Multidimensional classical and quantum cosmology
  with intersecting $p$-branes, hep-th/9708157;
  {\it J. Math. Phys.}, {\bf 39}, 2866-2889 (1998).

 \bibitem{Isbr}
  V.D. Ivashchuk, Composite S-brane solutions
  related to Toda-type systems, {\it Class. Quantum
  Grav.} {\bf 20}, 261-276 (2003); hep-th/0208101.

 \bibitem{IK}
  V.D. Ivashchuk and S.-W. Kim, Solutions with intersecting p-branes
  related to Toda chains, {\it J. Math. Phys.}, {\bf 41} (1)
  444-460 (2000); hep-th/9907019.

 \bibitem{IMtop}
  V.D. Ivashchuk and V.N. Melnikov,  Exact solutions in
  multidimensional gravity with antisymmetric forms,
  topical review, {\it Class. Quantum Grav.} {\bf 18},
  R82-R157 (2001); hep-th/0110274.

 \bibitem{IM11}
 V.D. Ivashchuk and V.N. Melnikov,
 Intersecting p-Brane Solutions in Multidimensional Gravity and M-Theory,
 hep-th/9612089; {\it Grav. and Cosmol.} {\bf 2}, No 4, 297-305 (1996);
 Generalized Intersecting P-brane Solutions from
 Sigma-model, {\it Phys. Lett.}  {\bf B 403}, 23-30 (1997).

 \bibitem{AR}
 I.Ya. Aref'eva and O.A. Rytchkov,
 Incidence Matrix Description of Intersecting p-brane Solutions,
 hep-th/9612236.

 \bibitem{AIV}
 I.Ya. Aref'eva, M.G. Ivanov and I.V. Volovich,
 Non-extremal intersecting p-branes in various dimensions, hep-th/9702079;
 {\it Phys. Lett. } {\bf B 406}, 44-48 (1997).

 \bibitem{IMC}
 V.D. Ivashchuk and V.N. Melnikov,
 Sigma-model for the Generalized  Composite p-branes,
 hep-th/9705036; {\it Class. Quantum Grav.} {\bf 14}, 3001-3029 (1997);
 Corrigenda {\it ibid.} {\bf 15 } (12), 3941 (1998).



\bibitem{I}
V.D. Ivashchuk, Multidimensional Cosmology and Toda-like Systems,
{\it Phys. Lett. },  {\bf A 170},  16-20 (1992).

\bibitem{BZ0}
U. Bleyer and  A. Zhuk, Kasner-like, inflationary and
steady-state solutions in multidimensional cosmology,
{\it Astron. Nachrichten}, {\bf 317}, 161-173 (1996).

\bibitem{BZ1}
U. Bleyer and  A. Zhuk, Multidimensional integrable cosmological
models with negative external  curvature, {\it Grav. Cosmol.},
{\bf 1}, 106-118 (1995).

\bibitem{BZ2}
U. Bleyer and  A. Zhuk, Multidimensional integrable cosmological
models with positive external space curvature, {\it Grav.
Cosmol.} {\bf 1}, 37-45 (1995).

\bibitem{IM10}
V.D. Ivashchuk and V.N.Melnikov,
Multidimensional classical and quantum cosmology with
perfect fluid, {\it Grav. Cosmol.}, {\bf 1},   133-148 (1995);
hep-th/9503223.

\bibitem{IMZ}
V.D. Ivashchuk, V.N. Melnikov and A.I. Zhuk,
On Wheeler-DeWitt  Equation in Multidimensional Cosmology,
{\it Nuovo Cimento } {\bf B 104}, 575  (1989).



\end{thebibliography}
\end{document}